\documentstyle[11pt,aaspp4]{article}
\begin{document}
%\draft
\title{$\gamma$-ray burst after-glow: Confirming the cosmological 
fireball model}
\author{Eli Waxman}
\affil{Institute for Advanced Study, Princeton, NJ 08540; E-mail:
waxman@sns.ias.edu}
%\date{\today}
\begin{abstract}

The recent detection of delayed X-ray and optical
emission, ``afterglow,'' associated with $\gamma$-ray bursts (GRBs)
supports models, where the bursts are produced by 
relativistic expanding blastwaves, ``fireballs,'' at 
cosmological distances.
The detection of absorption lines in the optical afterglow 
of the GRB of 8 May 1997 confirms that the sources lie at cosmological 
distance. We show here that the new features detected in GRB970508 afterglow,
radio emission one week following the burst and a 2 day increase in 
optical flux, are consistent with the blastwave
model. The fireball optical depth at radio frequencies is much smaller than
previously estimated, which accounts for the observed radio emission.
The initial suppression of optical flux is consistent
with that predicted due to electron cooling.
The combined radio and optical data imply that the fireball energy
is $\sim10^{52}$erg, and that the density of the medium into which the 
blastwave expands is $\sim1{\rm cm}^{-3}$, 
a value typical for gas within galaxies. We predict the time
dependence of the radio flux and the absorption frequency, 
which constitute tests of
the fireball model as described in this paper.

\end{abstract}

\section{Introduction}

The origin of GRBs, bursts of 0.1---1MeV photons lasting for a few seconds, 
 remained unknown for over 20 years (\cite{Fishman}), primarily
 because GRBs were not  
detected until this year at wavebands other than $\gamma$-rays. 
Phenomenological considerations based on
$\gamma$-ray data were used to argue that the bursts
are produced by relativistic expanding blastwaves at cosmological distances
(\cite{Bohdan1,Goodman,RM1,Piran}). The fireball model, which predicts delayed
emission at wavelengths longer than $\gamma$-rays
(\cite{PRAG,KAG,RMAG,VAG}), has gained support (\cite{AGW,AGWMR}) from the
 recent  
detection (\cite{Paradijs}) of delayed X-ray and optical
emission associated with GRBs. The key new development is the
 availability from the BeppoSAX satellite of accurate positions for GRBs 
shortly after their detection. 

GRB970508 was detected by the BeppoSAX satellite (\cite{SAX}) 
on 8 May 1997. The
burst lasted for $\sim15$s, with $\gamma$-ray fluence $\sim3\times10^{-6}
{\rm erg\,cm}^{-2}$ carried mainly by photons of energy $\sim0.5$MeV
(\cite{BATSE}). 
Following the detection in $\gamma$-rays, X-ray (\cite{X}), 
optical (\cite{O6654,O6655,O6658,O6660,O6661,O6663}), and radio (\cite{R})
emission varying on time scale of days was observed from the direction
of the GRB. The location of the GRB is determined with accuracy better than 
$\sim3\arcmin$, while the locations of the X-ray, optical and radio
sources are determined to better than $50\arcsec$, $1\arcsec$, and 1mas  
respectively. The fact 
that the locations of all sources coincide, and the unusual variability 
observed following the GRB, strongly suggest that the X-ray, optical 
and radio emission are associated with the object producing the GRB.
A spectrum of the optical transient, taken 2 days after the GRB,
shows a set of absorption features, 
associated with Fe II and MgII and shifted to long wavelength,
implying that the absorbing system lies at a cosmological redshift z=0.835
(\cite{absorption}). This sets a lower limit to the GRB redshift, $z\ge0.835$,
and to the energy emitted by the source in $\gamma$-rays (assuming isotropic
emission), $E_\gamma\ge10^{51}$erg. Intensive monitoring of
GRB970508 revealed two new afterglow features: radio emission was observed one
week following the burst (\cite{R}), and the optical flux was observed to 
increase for 2 days following the burst (\cite{O6655,O6658}). 

The early detection of radio emission, combined with optical data, 
appears to be inconsistent with published 
fireball model predictions (\cite{PRAG,RMAG}). We show here, however, that the 
fireball optical depth at radio frequency is significantly smaller than
previously estimated, and that the same process responsible for the optical
afterglow may also produce the observed radio emission. The combined radio
and optical data are shown to be consistent with the fireball model, and to 
provide information on the fireball parameters and on the ambient medium
into which the blastwave expands. The suppression of optical flux 
at early times provides evidence for the predicted 
suppression due to electron cooling (\cite{AGW}), 
and suggests that the GeV emission observed for several hours 
following several strong GRBs (\cite{GeV}) is produced by 
inverse-Compton scattering of afterglow X-ray photons. 

Our goal is to present a qualitative analysis of the new afterglow features  
observed in GRB970508, in order to identify the physical processes that
are responsible for the observed behavior. We therefore adopt a simple
approximate description of the fireball expansion, that permits the
derivation of 
the main results without complicated calculations. Our values for numerical
parameters should be considered as order of magnitude estimates.

\section{The fireball model and predicted X-ray, optical and radio afterglow}

The underlying source, which produces the initial explosion that drives
the expanding fireball is unknown (although several plausible candidates
have been proposed, see, e.g., \cite{Piran} for review). 
However, the $\gamma$-ray observations suggest the following
scenario for the emission of the observed $\gamma$-rays
(\cite{Bohdan1,Goodman,RM1}). A compact, $r_0\sim10^7$cm,
source releases an energy $E$ comparable to that observed in $\gamma$-rays,
$E\sim10^{51}$erg, over a time $T<100$s. The large energy density in the 
source results in an optically thick plasma that expands and accelerates to
relativistic velocity. After an initial acceleration phase, the fireball
energy is converted to proton kinetic energy. A cold shell of 
thickness $cT$ is formed and continues to expand with time independent
Lorentz factor $\gamma\sim300$. The GRB is produced once the kinetic energy
is dissipated at large radius, $r>10^{13}$cm, due to internal collisions 
within the ejecta (\cite{Xu,RnM}) and radiated as $\gamma$-rays
through synchrotron and possibly inverse-Compton emission of shock accelerated 
electrons. 
 
Following internal collisions, which
convert part of the energy to radiation and which result
from variations in $\gamma$ across the expanding shell, 
the fireball rapidly cools and
continues to expand with approximately uniform Lorentz factor $\gamma$.
As the cold shell expands it drives a relativistic shock (blastwave)
into the surrounding
gas, e.g. into the inter-stellar medium (ISM) gas if the explosion 
occurs within a galaxy. In what follows, we refer to the surrounding gas
as ``ISM gas,'' although the gas need not necessarily be inter-stellar.
The shock propagates with Lorentz factor $\gamma_s=
2^{1/2}\gamma$, and behind it the (rest frame) number and energy densities
of the shock heated ISM are $n'=4\gamma n$ and $e'=4\gamma^2nm_pc^2$,  
respectively, where $n$ is the ISM number
density ahead of the shock. The width of the shock heated ISM shell
is $r/4\gamma^2$, where $r$ is the fireball radius.

At late times, the blast wave approaches a self similar behavior. The expansion
is well approximated by the self-similar solution 
for radii $r>r_c$, where (\cite{AGW})
\begin{equation}
r_c=\left({ET\over4\pi nm_pc}\right)^{1/4}=
2\times10^{16}\left({E_{52}T_{10}\over n_1}\right)^{1/4}{\rm cm}.
\label{rc}
\end{equation}
Here the fireball energy is $E=10^{52}E_{52}$erg, the ISM density is 
$n=n_1{\rm cm}^{-3}$, and $T=10T_{10}$s. We have chosen to give our numeric
results using a fireball energy value that is somewhat higher than the typical
energy observed in $\gamma$-rays, $E_\gamma\sim10^{51}{\rm erg}$, since the
conversion of fireball energy to $\gamma$-rays is not expected to be $100\%$
efficient. For $r\ge r_c$, the shell Lorentz factor is
\begin{equation}
\gamma=\left({r_c\over2Tc}\right)^{1/2}(r/r_c)^{-3/2}=
190\left({E_{52}\over T_{10}^3 n_1}\right)^{1/8}(r/r_c)^{-3/2}.
\label{gt}
\end{equation}
Photons emitted from the shell at radius $r$ are seen by a distant
observer at a time $t=r/2\gamma^2c=T(r/r_c)^4$ after the GRB, with
arrival time spread comparable to $t$. The main contribution to the time
delay and spread is due to two effects. First, the radiation seen by a distant 
observer is emitted from a cone of the fireball around the source-observer line
of sight, with an opening angle $\sim1/\gamma$. Photons emitted from
such a cone are spread over $t=r/2\gamma^2c$. Second, as we show below,
during most of the afterglow the synchrotron cooling time of electrons is 
larger than the fireball expansion time, $\sim r/\gamma c$ 
as measured in the fireball frame. Thus, electrons emit radiation over a time
$\sim r/\gamma c$ in the fireball frame, corresponding to a time 
$\sim r/2\gamma^2c$ as seen by a distant observer (Note, that the delay due
to the difference $t_s$ between the shock propagation time to radius $r$ and 
$r/c$, $t_s=r/16\gamma^2c$, is not the main factor determining the 
time $t$ at which radiation from radius $r$ is observed).

The shock driven into the ISM continuously heats new gas, and
produces relativistic 
electrons that may produce the delayed radiation observed on time scales of
days to months. In order
to calculate the synchrotron emission from the heated ISM shell we need
to determine the magnetic field and electron energy within the heated shell. 
We assume that the
magnetic field energy density (in the shell rest frame) is a fraction 
$\xi_B$ of the
equipartition value, $B^2/8\pi=\xi_B e'$, and that the electrons
carry a fraction $\xi_e$ of the energy. Since the Lorentz factor associated
with the thermal motion of protons in the shell rest frame is $\gamma$,
this implies that the Lorentz factor of the random motion of a typical
electron in the shell rest frame is $\gamma_{em}=\xi_e\gamma m_p/m_e$. 
With these
assumptions, and using eqs. (\ref{rc}) and (\ref{gt}), 
we
find that the observed frequency of synchrotron emission from typical
electrons, $\nu_m\simeq\gamma\gamma_{em}^2eB'/2\pi m_ec$, is
\begin{equation}
\nu_m=2\times10^{14}\left({1+z\over2}\right)^{1/2}
(\xi_e/0.2)^2(\xi_B/0.1)^{1/2}E_{52}^{1/2}t_{\rm day}^{-3/2}{\rm Hz},
\label{nu}
\end{equation}
where $t=1t_{\rm day}$days, and $z$ is the cosmological 
redshift of the burst. We have chosen to give numerical results using $\xi_e
\sim\xi_B\sim0.1$, since such values are typically required for the production
of the GRB itself.
We show below that the electron synchrotron cooling time is longer than the
dynamical time, $t_d=r/\gamma c$, the time for significant fireball expansion.
In this case, the observed intensity at $\nu_m$ is  (\cite{AGW})
\begin{equation}
F_{\nu_m}=1\,\left({1 + z\over 2}\right)^{-1}\left[{1 - 1/\sqrt{2}\over
1 - 1/\sqrt{1 + z}}\right]^{2}
n_1^{1/2}(\xi_B/0.1)^{1/2}E_{52} {\rm mJy}.
\label{Fm}
\end{equation}
Here (and throughout the paper) we assume a flat universe with Hubble constant 
$H_0=75{\rm km/s\,Mpc}$. The Jy flux unit is $1{\rm Jy}=10^{-23}
{\rm erg/cm^2\,s\,Hz}$.

A natural prediction
of the fireball model, see eqs. (\ref{nu}) and (\ref{Fm}), is   
optical emission at a level of 1mJy
at a delay of order 1day following the GRB. 

Emission at $\nu>\nu_m$ is produced by electrons with Lorentz factor
higher 
than $\gamma_{em}$. If the electron distribution follows a power-law,
$dN_e/d\gamma_e\propto
\gamma_e^{-p}$ for $\gamma_e>\gamma_{em}$, as expected for shock acceleration,
then for $\nu>\nu_m$ 
%(defining $\alpha\equiv(p-1)/2$):
\begin{equation}
F_\nu=F_{\nu_m}[\nu/\nu_m(t)]^{-\alpha},
\label{Ft}
\end{equation}
with $\alpha=(p-1)/2$. Although $p$ is expected to be similar for the GRB and 
for the afterglow,
$\alpha$ during the after-glow is expected to be smaller by $1/2$ due to 
increase in cooling time: 
the synchrotron cooling time must be short
compared to the dynamical time during the GRB, resulting in $\alpha=p/2$,
while it is long during the afterglow, giving $\alpha=(p-1)/2$ (\cite{AGW}). 
For the GRB $\alpha\sim1$, implying $\alpha\sim0.5$
for the after-glow. 

The synchrotron emission of electrons with Lorentz
factor $\gamma_{em}$ is concentrated mainly at frequencies $\nu\sim\nu_m$.
However, the emission extends to lower frequencies, with power radiated
per unit frequency proportional to $(\nu/\nu_m)^{1/3}$. Thus,
we expect the synchrotron flux to
extend to $\nu<\nu_m$ following eq. (\ref{Ft}) with $\alpha=-1/3$. This
implies that the flux at a fixed frequency $\nu$ increases with time as 
$t^{1/2}$ as long as $\nu<\nu_m$.
We have so far assumed that the fireball optical depth to synchrotron 
absorption is small. This may not be the case for low frequencies. The
synchrotron optical depth at $\nu_m$ is small, $\tau_m=10^{-11}
\xi_e^{-5}\xi_B^{-1/2}E_{51}^{-1/2}n_1t_{\rm day}^{5/2}$. For $\nu>\nu_m$,
$\tau\propto\nu^{-(p+4)/2}$, while $\tau\propto\nu^{-5/3}$ for $\nu<\nu_m$
due to the $\nu^{1/3}$ low frequency synchrotron tail. 
The frequency for which the optical depth is 1, 
$\nu_A=\nu_m\tau_m^{3/5}$, is
\begin{equation}
\nu_A=1\left({1+z\over2}\right)^{-1}(\xi_e/0.2)^{-1}(\xi_B/0.1)^{1/5}
E_{52}^{1/5}n_1^{3/5}{\rm\, GHz},
\label{nuA}
\end{equation}
and the flux at $\nu_A<\nu\ll\nu_m$, 
$F_A=2F_{\nu_m}(\nu/\nu_m)^{1/3}$, is
\begin{equation}
F_\nu=0.3\left({1+z\over2}\right)^{-7/6}\left[1-1/\sqrt 2\over 
1-1/\sqrt{1+z}\right]^2(\xi_e/0.2)^{-2/3}(\xi_B/0.1)^{1/3}E_{52}^{5/6}
n_1^{1/2}(t/1{\rm week})^{1/2}(\nu/10{\rm GHz})^{1/3}{\,\rm mJy}.
%F_A=0.2\left({1+z\over2}\right)^{-3/2}\left[1-1/\sqrt 2\over 
%1-1/\sqrt{1+z}\right]^2(\xi_e/0.2)^{-1}(\xi_B/0.1)^{4/10}E_{52}^{9/10}
%n_1^{7/10}(t/1{\rm week})^{1/2}{\,\rm mJy}.
\label{FA}
\end{equation}
Eq. (\ref{nuA}) and (\ref{FA}) indicate, that radio emission at a level of
1mJy is expected at the $\sim10$GHz range on time scale of weeks.
The emission should be suppressed below $\sim1$GHz due to high optical 
depth. 

We note here, that a population of electrons with $\gamma_e\ll\gamma_{em}$ may
also contribute to the flux and optical depth at $\nu\ll\nu_m$. Although in 
the model described in this paper 
the electron distribution is dominated by electrons
with Lorentz factor $\gamma_e\sim\gamma_{em}$, the distribution may extend
to low energy, $\gamma_e\ll\gamma_{em}$, 
without significantly affecting the results (\ref{nu}--\ref{Ft}), provided
that low energy electrons constitute a small fraction of the electron 
population. The distribution may extend, e.g., to $\gamma_e\ll\gamma_{em}$ as 
$dN_e/d\gamma_e\propto\gamma_e^{-p'}$ with $0\le p'\ll1$ or $p'<0$
(for $p'\gtrsim1$ 
the fraction of electrons with $\gamma_e\sim\gamma_{em}$ is small). 
While the contribution of such low energy electron population to the flux
at $\nu<\nu_m$, $F_\nu\propto\nu^{(1-p')/2}$, is not large compared to that
given in (\ref{FA}), their contribution to the optical depth at $\nu<\nu_m$,
$\tau\propto\nu^{-(p'+4)/2}$, may result in self-absorption frequency which
is significantly higher than (\ref{nuA}). For $p'=0$ we have
$\nu_A=\nu_m\tau_m^{1/2}$, i.e.
\begin{equation}
\nu_A=4\left({1+z\over2}\right)^{-3/4}(\xi_e/0.2)^{-1/2}(\xi_B/0.1)^{1/4}
E_{52}^{1/4}n_1^{1/2}(t/1{\rm week})^{-1/4}{\rm\, GHz}.
\label{nuAp}
\end{equation}
Eq. (\ref{nuAp}) implies, that 
the existence of low energy electron population does not change the conclusion
that radio emission at a level of
1mJy is expected at the $\sim10$GHz range on time scale of weeks. 
(Our estimate for $\nu_A$ is significantly lower than previous estimates [e.g.
\cite{PRAG,RMAG}] since we have taken into account the fact that the electron
distribution flattens below $\gamma_{em}$.)

\section{GRB970508}

Let us now compare the fireball model predictions with observations of
GRB970508.
The detected absorption lines imply that the GRB source redshift is 
$z\ge0.835$, and the absence of Lyman-$\alpha$ lines imply $z<2.1$ 
(\cite{absorption}). The absence of CIV absorption further implies $z<1.6$
(N. Arav and D. Hogg, private communication). We therefore adopt $z=1$ 
as the GRB source redshift. 

The radio flux detected 6 days after the GRB
is well described by $0.2(\nu/3{\rm GHz}){\rm mJy}$
in the range of 1 to 10GHz (\cite{R}). The $\nu^1$ dependence, which is steeper
than $\nu^{1/3}$ expected from the low frequency tail of synchrotron
emission in the absence of absorption, implies that the radio emission
is partly absorbed, with stronger absorption at lower frequencies due to
increase in optical depth. In our model, $F_\nu\propto
\nu^{1/3}$ for $\nu/\nu_A\gg1$, and $F_\nu\propto\nu^2$ 
for $\nu/\nu_A\ll1$ (in the presence of low energy 
electrons, $dN_e/d\gamma_e\propto\gamma_e^{-p'}$ with $p'>-2/3$ for 
$\gamma_e\ll\gamma_{em}$, the dependence of flux on frequency is stronger,
$F_\nu\propto\nu^x$ with $x>2$ for $\nu/\nu_A\ll1$). 
Radio observations therefore indicate that $1{\rm GHz}<\nu_A<10$GHz.
The observed self-absorption frequency and the level of detected radio flux 
are in agreement with model predictions (\ref{nuA}--\ref{nuAp}). 
The upper limit of $7{\rm mJy}$ 
at 6 day delay at $90$GHz (\cite{R1}) is also 
consistent with the model $\nu^{1/3}$
dependence at $\nu>\nu_A$, which predicts a flux at 90GHz
of $1{\rm mJy}$.
At delays shorter than 6 days, only upper limits to the flux at 1.4GHz are
available. The upper limits are comparable to the flux detected at 6day delay,
and therefore, while being consistent with the model, 
do not allow to establish the transient nature of the radio
emission or test the model prediction, that the flux should be increasing 
with time. 

The R band, $\nu_R=4\times 10^{14}$Hz, observations at 2.0 to 5.5 day delay
(\cite{O6655,O6658,O6660,O6661,O6663}) are well described by a power-law,
$F_R=37\mu{\rm Jy}(t/2{\rm day})^{-1.3}$. 
The power law behavior is consistent with the model prediction
(\ref{Ft}). The approximately
constant flux during 1 to 2 days delay indicate that the peak of the 
synchrotron emission $\nu_m$ passed through the R band at a delay
$t_R\sim1$day. This is consistent with the model prediction (\ref{nu}).
(Note that extrapolation of the $t^{-1.3}$ behavior to $t=1$day
results in a flux $\sim2$ times higher than observed. However, since
we do not expect
a sharp peak with in the flux observed at a given frequency as $\nu_m$ drops
below this frequency, due to the spread in photon arrival times, the 
approximately constant flux observed near the peak on time scale comparable
to the delay is consistent with the model). 
The normalization of the power-law fit to the R
band observations is $\sim3$ times lower than implied by eq. (\ref{Fm}),
(\ref{nu}), and
(\ref{Ft}). However, given the simple description of the
fireball behavior adopted in this paper, one should not draw conclusions 
based on this numerical discrepancy. The $t^{-1.3}$ decline of R band flux
implies, through (\ref{Ft}), $\alpha\simeq0.8$, or
$p=2.6$. The implied frequency dependence of the flux is consistent with
observations at other optical bands, although the narrow frequency range does
not allow accurate determination of $\alpha$ from the frequency dependence.

Consider next the optical (\cite{O6654,O6655}) detection at $\sim7$hr delay,
and the X-ray detection at $10$hr delay (\cite{X}). At early time, $t<1$day,
the R band frequency is smaller than $\nu_m$, the frequency of radiation
emitted by the typical fireball electrons, and the $R$ band flux should 
increase as $t^{1/2}$. Using the the power-law fit to the observed 
R-band flux for $t>1$day, and a $t^{1/2}$ scaling for $t<1$day, the predicted
flux at 7hr delay is $50\mu{\rm Jy}$, significantly above the observed flux,
$11.1\pm2.2\mu{\rm Jy}$. 
This discrepancy may indicate the detection of the
predicted (\cite{AGW}) suppression of optical
flux at early times due to
rapid electron cooling. The ratio of synchrotron cooling time, 
$t_s=6\pi m_ec/\sigma_T\gamma_eB'^2$, to fireball deceleration time, 
$t_d\sim\gamma/\dot{\gamma}\gamma=2r/3\gamma c$, is 
$t_s/t_d\sim2(\xi_e/0.2)^{-1}(\xi_B/0.1)^{-1}E_{52}^{-1/2}
t_{\rm day}^{1/2}$. Thus, our assumption that the synchrotron cooling time is 
longer than the dynamical time is valid for $t\gtrsim1$day. At 7hr delay 
the cooling
time is comparable to the deceleration time, and the electrons
cool on a dynamical time scale. At this time, synchrotron emission is
significantly suppressed due to inverse-Compton emission (\cite{AGW}), by
which electrons lose a significant fraction of their energy to the production
of photons of high energy, exceeding 1~GeV during the first hours
following the burst. This may account for the delayed GeV
emission observed in several strong bursts. 
Finally, the X-ray emission detected at $\sim10$hr delay 
(\cite{X}), with a flux of $\sim 0.03 \mu{\rm Jy}$ at $10^{18}$Hz, 
is consistent with 
the above model when the suppression due to electron cooling is taken into
account.

\section{Conclusions}

The fireball model for GRBs is consistent with afterglow
observation in X-ray, optical and radio wave bands, provided the kinetic
fireball energy is $E\sim10^{52}$erg, the density into which the blastwave
expands is $\sim1{\rm cm}^{-3}$, and provided that the fraction of energy
carried by electrons and magnetic field is $\xi_e\sim\xi_B\sim 0.1$.
The inferred kinetic energy is consistent with the
observed $\gamma$-ray energy, $E_\gamma\sim10^{51}$erg, and implies that
the efficiency with which kinetic energy is converted to $\gamma$-rays
is of order $10\%$. The density of the ambient medium is typical for  
interstellar gas, therefore suggesting that the explosions take place within
galaxies. The requirement for significant energy to be carried by electrons
and magnetic field is consistent with what is usually assumed for the 
production of the GRB itself.
We have implicitly assumed that the fireball is spherical. However, our
analysis is valid also for the case that the fireball is a jet,
as long as the jet opening angle is larger than $1/\gamma$. The
fireball Lorentz factor during the radio observations is not very large,
$\gamma=4(E_{52}/n_1)^{1/8}(t/1{\rm week})^{-3/8}$. This implies that the
opening angle of the jet is not very small.

Our model predicts that the frequency, at which the optical depth for
synchrotron absorption is unity, should change slowly with time,
$\nu_A\propto t^{-x}$ with $0\le x\le 1/4$, and that on time scale of weeks
$1{\rm GHz}<\nu_A<10{\rm GHz}$ (cf. eq. [\ref{nuA}--\ref{nuAp}]). 
The radio flux at $\nu>\nu_A$ should rise as $t^{1/2}$ following the GRB. 
These predictions can be tested with more
frequent radio observations following future GRBs. (For GRB970508 the observed
self-absorption frequency and radio flux at 6day delay are consistent with 
model predictions. However, at shorter delays only upper limits are available,
which do not allow to establish the transient nature of the source or test
the above predictions).
Since the Lorentz factor after one week is not large, deviations from
the simple scaling laws derived here may be observed at 
later times, due to the deceleration of the fireball to 
velocity which is not highly relativistic.

\paragraph*{Acknowledgments.} 

I thank J. N. Bahcall,
B. Paczy\'nski and R. Sari for helpful discussions. This research
was partially supported by a W. M. Keck Foundation grant 
and NSF grant PHY95-13835.

\end{document}